\documentstyle{mn}

\title[THUMPER on JCMT]{First ground-based
200-$\mu$m observing with THUMPER on JCMT -- 
sky characterisation and planet maps}

\author[D. Ward-Thompson et al.]
{D. Ward-Thompson$^1$, P. A. R. Ade$^1$, H. Araujo$^2$, I. Coulson$^3$, 
J. Cox$^1$, G. R. Davis$^3$, \cr
Rh. Evans$^{1,4}$, M. J. Griffin$^1$, W. K. Gear$^1$, P. Hargrave$^1$, 
P. Hargreaves$^1$, \cr
D. Hayton$^1$, B. J. Kiernan$^1$, S. J. Leeks$^{1,5}$, P. Mauskopf$^1$, 
D. Naylor$^6$, N. Potter$^1$, \cr
S. A. Rinehart$^7$, R. Sudiwala$^1$, C. R. Tucker$^1$, R. J. Walker$^{1,2}$,
S. L. Watkin$^1$ \\
$^1$ Department of Physics and Astronomy, Cardiff University,
5 The Parade, Cardiff, CF24 3YB \\
$^2$Blackett Laboratory, Imperial College, London SW7 2BW \\
$^3$Joint Astronomy Centre, 665 N. A'Ohoku Street, Hilo, Hawaii, USA \\
$^4$Glamorgan University, Pontypridd, CF37 1DL \\
$^5$European Space Astronomy Centre, Villafranca del Castillo,
PO Box 50727, 28080 Madrid, Spain \\
$^6$Department of Physics, University of Lethbridge, 4401 University Drive, 
Lethbridge, Alberta, Canada \\
$^7$Laboratory for Astronomy and Solar Physics, Goddard Space Flight Center,
Washington D.C., USA}

\date{Accepted 2005 September 1; received 2005 September 1; in original form
2005 July 1.}

\pagerange{000--000}

\begin{document}

\label{firstpage}

\maketitle

\begin{abstract}
We present observations that were
carried out with the Two HUndred Micron PhotometER (THUMPER) mounted on
the James Clerk Maxwell Telescope (JCMT) in Hawaii, at a
wavelength of 200~$\mu$m (frequency 1.5~THz).
The observations utilise a small atmospheric
window that opens up at this wavelength under very dry conditions at
high-altitude observing sites. The atmosphere was calibrated using the 
sky-dipping method and a relation was established
between the optical depth, $\tau$, at 1.5~THz and that at 225~GHz:
$\tau_{1.5{\rm THz}} = (95 \pm 10) \times \tau_{225{\rm GHz}}$.
Mars and Jupiter were mapped from the ground at this wavelength for the
first time, and the system characteristics measured. A noise equivalent
flux density (NEFD) of $\sim$65$\pm$10~Jy (1$\sigma$ 1s)
was measured for the
THUMPER--JCMT combination, consistent with predictions based upon
our laboratory measurements.
The main-beam resolution of 14~arcsec was confirmed
and an extended error-beam detected at roughly two-thirds of
the magnitude of the main beam. Measurements of the Sun allow us
to estimate that the fraction of the power in the main beam is
$\sim$15\%, consistent with predictions based on modelling the dish
surface accuracy. It is therefore
shown that the sky over Mauna Kea is suitable for astronomy at
this wavelength under the best conditions. However, higher or drier sites
should have a larger number of useable nights per year.
\end{abstract}

\section{Introduction}

Far-infrared astronomy is largely the preserve of telescopes borne on 
satellites, balloons and aircraft. The Infra-Red Astronomical Satellite,
IRAS (Beichman et al., 1988) operated at wavelengths as long as 100~$\mu$m.
The Kuiper Airborne Observatory, KAO 
(Cameron, Bader \& Mobley 1971; Harvey 1979), the Infrared
Space Observatory, ISO (Kessler et al., 1996),
and the Spitzer Space Telescope (Werner et al., 2004)
were designed to operate out to
200~$\mu$m. A number of planned missions such as Herschel and Planck aim 
to cover still longer wavelengths. However,
a problem with all such airborne and satellite telescopes is that they 
are unlikely to
be as large as telescopes that can be built on the ground.
Consequently, airborne and orbiting telescopes do not have the angular
resolution of comparable ground-based instruments. Therefore, any 
instruments that can be operated from the ground at these wavelengths
and mounted on large telescopes
will automatically have a resolution advantage.

Only a limited number of previous ground-based astronomical 
observations have been made at far-infrared wavelengths.
This is because the atmosphere is largely opaque
in this regime. However, spectral measurements of the atmosphere,
coupled with detailed modelling of atmospheric properties, have shown that 
there are some small windows that do open up across this range on very
high, dry sites under good conditions. One such window is that at 200~$\mu$m,
corresponding to a frequency of 1.5~THz. 

\begin{figure}
\setlength{\unitlength}{1mm}
\noindent
\begin{picture}(70,60)
\put(-16,-43){\includegraphics{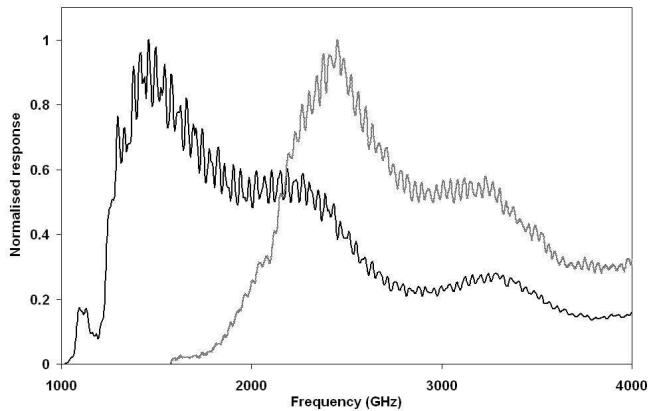}}
\end{picture}
\caption{Plot of normalised
crystal response versus frequency for an unstressed
crystal (lighter line to the right) and a stressed crystal (heavier
line to the left) in the THUMPER detector array. Note how the peak
response has moved to the required THUMPER frequency of 1500~GHz.}
\label{crystals}
\end{figure}

Previous ground-based measurements at this frequency have only been made
spectroscopically. No previous ground-based imaging camera has been built
to work at this frequency. The spectroscopic observations include
Fourier Transform Spectrometer (FTS) measurements at Pampa la Bola in Chile
(Matsuo, Sakamoto \& Matsushita 1998; Matsushita et al., 1999).
These latter authors derived measurements relating the optical depths
in different atmospheric pass-bands, including that at 200~$\mu$m.
Ground-based FTS measurements at Mauna Kea in Hawaii have been carried
out by a number of groups (e.g. Serabyn \& Weisstein 1996; Serabyn et al.,
1998; Pardo et al., 2004). All have confirmed the existence of an
atmospheric window at 200~$\mu$m, albeit with varying levels of transmission,
depending upon the conditions.

Paine et al., (2000) carried out similar FTS measurements at both Mauna Kea
in Hawaii and at Chajnantor in Chile and measured the atmospheric 
transmission from about 300 to 3000~GHz.
They measured the 1.5-THz window and found significant
transmission in the wave-band of interest at both locations.
Other work at this frequency has included plans for
a heterodyne system to operate at the South Pole by a group who have
made FTS sky measurements, broadly confirming the above-mentioned results
(Chamberlin et al., 2003; Gerecht et al., 2003).

In this paper, we present the results from the
first commissioning run of the Two HUndred Micron PhotometER (THUMPER) at
the James Clerk Maxwell Telescope (JCMT). This instrument is a prototype
camera built at Cardiff University to demonstrate
the viability of carrying out
high-resolution broad-band imaging at 200~$\mu$m from the ground.

\begin{figure}
\setlength{\unitlength}{1mm}
\noindent
\begin{picture}(70,60)
\put(-16,-43){\includegraphics{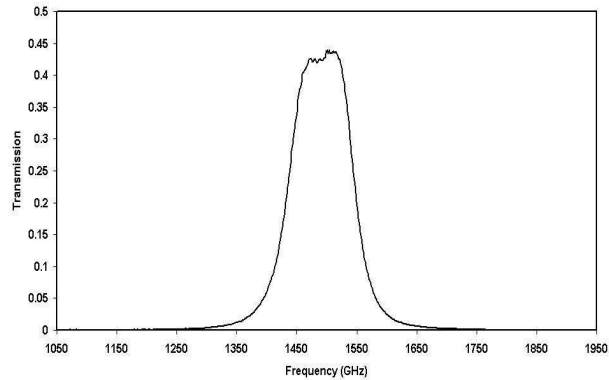}}
\end{picture}
\caption{Plot of filter transmission versus frequency for the complete
filter stack in THUMPER. The atmospheric central frequency corresponds to
1500~GHz.}
\label{filters}
\end{figure}

\section{The Instrument}

THUMPER consists of a 7-channel array of photoconductor pixel elements
arranged on a hexagonal grid 
(Ward-Thompson et al., 2002; Walker et al., 2003; Evans et al., 2005).
This is designed to match the layout of the central
pixels in the Sub-millimetre Common User Bolometer Array, SCUBA
(Holland et al., 1999), when placed at the Nasmyth focus of 
the James Clerk Maxwell Telescope (JCMT) on Mauna Kea, Hawaii. This is
at an altitude of 4092~m, and is one of the best sites for
sub-millimetre observations in the world.

The THUMPER detectors are photo-conductors made from gallium-doped germanium 
(in the ratio Ge:Ga $\sim$ 10$^5$:1)
crystal semi-conductors, which have 
been stressed to move their peak responsivities into the 200-$\mu$m
waveband (Walker et al., 2003; Walker 2004). Figure~\ref{crystals} shows
a typical crystal responsivity as a function of frequency. The lighter line
to the right
shows the natural crystal response. The heavier line
to the left shows the response
after the crystal has been mounted in the camera and stressed to its
optimal value. A series of
filters matches the band-pass of the instrument to the atmospheric window.
Figure~\ref{filters} shows the combined band-pass filtering of the system.

A drawing of one of the stressing blocks is shown in
Figure~3(a). The crystals are mounted in the integrating cavities
on the front of the stressing block (on the left of Figure~3(a)) and
a stressing rod is pushed up into the rear of the block 
(on the right of Figure~3(a))
using a stressing screw. This in turn operates a lever which applies
the stress downwards through the line of crystals by means of further
stressing rods.

\begin{figure*}
\setlength{\unitlength}{1mm}
\noindent
\begin{picture}(140,120)
\put(-46,0){\includegraphics{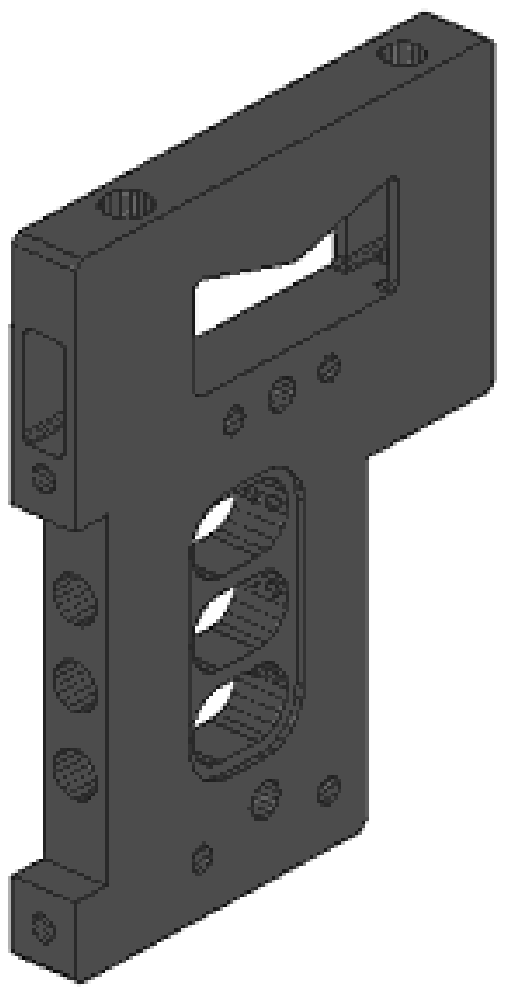}}
\put(46,0){\includegraphics{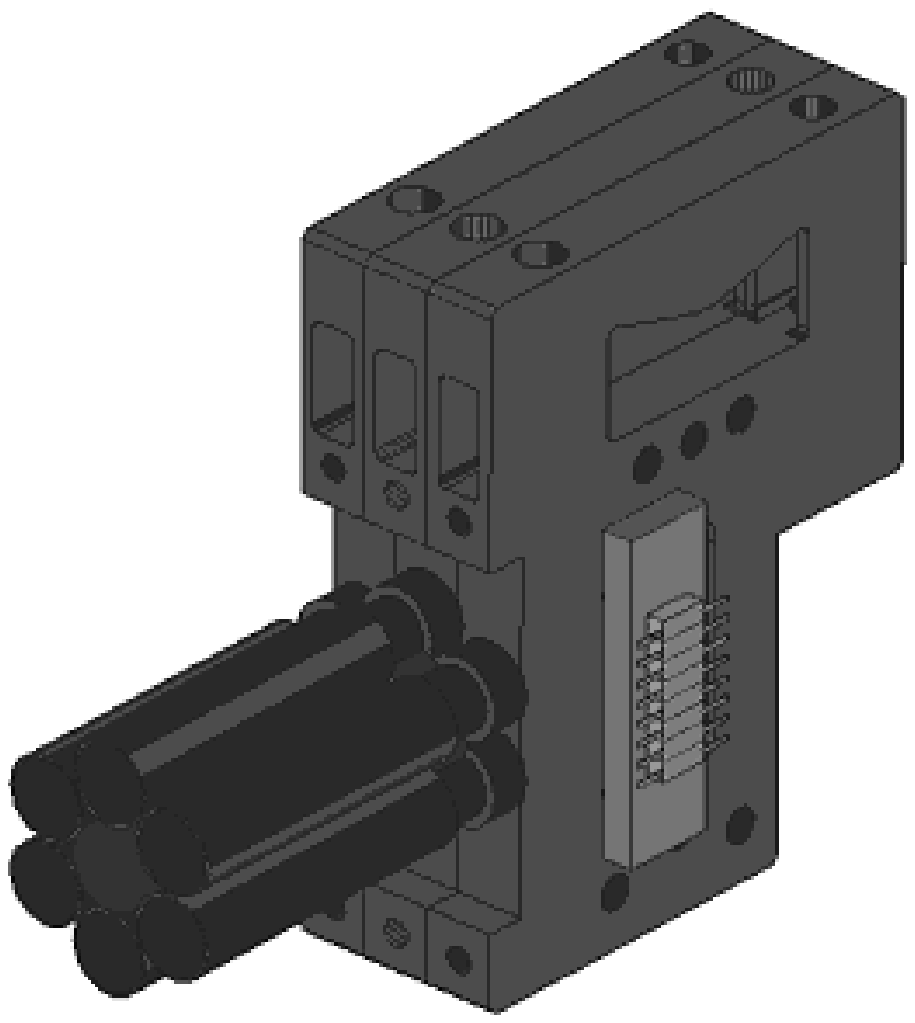}}
\put(-46,-63){\includegraphics{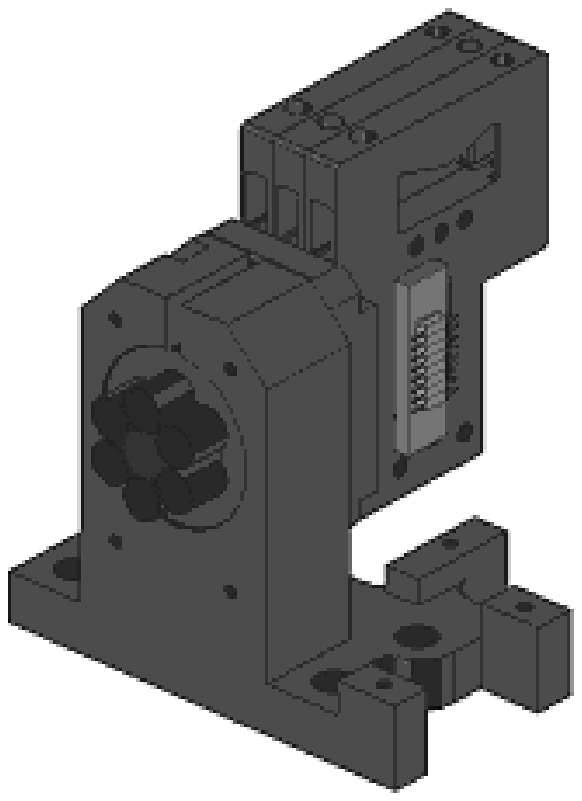}}
\put(46,-63){\includegraphics{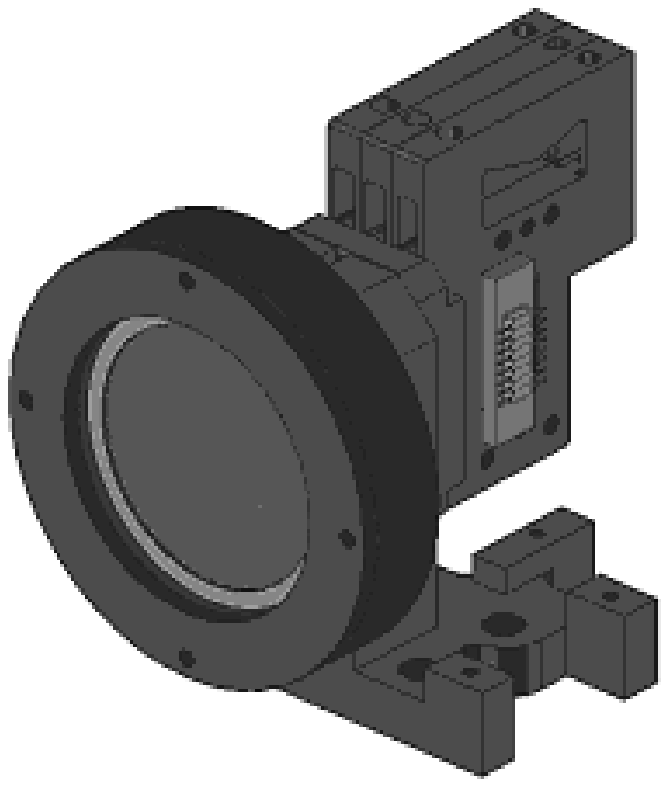}}
\end{picture}
\caption{The THUMPER focal plane array: (a: top left) a stressing block in
which the crystals are mounted; (b: top right) the three stressing blocks
fixed together to form the focal plane array, with feed-horns on the front
of the detectors; (c: lower left) the focal plane array mounted on its
stand; (d: lower right) the complete array and mount with the filter
stack mounted on the front.}
\label{fig3}
\end{figure*}

There are three stressing blocks. The centre block holds three crystals
and the two side blocks hold two crystals each, such that when the three
blocks are fixed together, as in Figure~3(b), the crystals make a hexagonal
arragement. The crystal integrating cavities
are coupled to the telescope by individual feed-horns,
as shown in Figure~3(b), with a resolution of
14 arcsec, as for the SCUBA 850-$\mu$m array (Holland et al., 1999). The
THUMPER focal plane array can also be adjusted to match the SCUBA 450-$\mu$m
array if desired, although this mode was not used on this occasion. 
We chose not to make the feed-horns diffraction-limited on the telescope
because the dish surface accuracy is too low to provide 
a significant fraction of the total incident power in a diffraction-limited
beam at this wavelength.

The focal plane array is mounted in a stand as shown in Figure~3(c) and
the filter stack attached to the front of the array as seen in Figure~3(d).
The whole array is
cooled to liquid Helium temperature (3.7~K at JCMT) in a long hold-time
cryostat, built by Queen Mary College Instruments
(now based at Cardiff University) and Thomas Keating Ltd. 
The system detector quantum efficiency (DQE) was measured in the laboratory.
Calculations were performed based upon the optical set-up of the
instrument at the telescope, including estimates of the dish surface 
accuracy. Incorporating all of these factors we estimated before
going to the telescope that the
noise equivalent flux density (NEFD) of the JCMT-THUMPER arrangement
should be $\sim$50--70~Jy (1$\sigma$~1s -- Walker 2004).

\section{Observations}

The cryostat was mounted on the right Nasmyth platform of the JCMT. Two
lenses, made of high-density polyethylene, were used to bring the beam
from the telescope to the f/5.75 focus required by THUMPER (Walker et al. 
2003). The instrument was commissioned during 
Director's Discretionary Time in two 4-night observing runs on
2005 March 20-23, and 2005 April 6-9. Optical alignment and instrument setup
were carried out prior to this. Data acquisition was carried out via a
stand-alone computer operating the LabVIEW software system
(Laboratory Virtual Instrument Engineering Workbench). Only
one of the scheduled nights was good enough for astronomical observations
at this wavelength, 2005 April 9, when our observing was scheduled 
from 01.30 to 09.30 Hawaiian Standard Time (UT 11.30 to 
19.30). Data from that night are presented in this paper.

\begin{figure}
\setlength{\unitlength}{1mm}
\noindent
\begin{picture}(70,60)
\put(-16,-43){\includegraphics{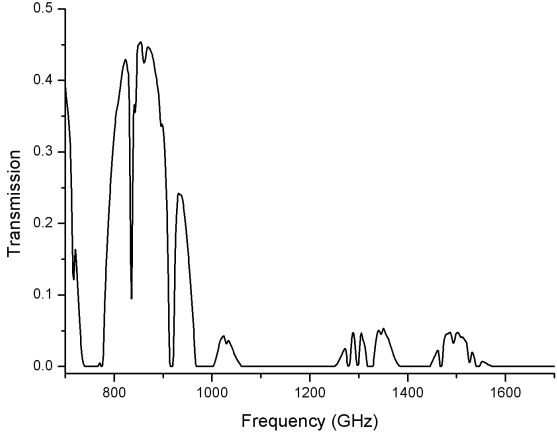}}
\end{picture}
\caption{Plot of atmospheric transmission versus frequency as we have
modelled, for 0.5mm PWV. Note the more frequently used sub-millimetre window 
at 850~GHz (350~$\mu$m). The 1.5-THz window is beginning to open up
in these conditions.}
\label{atmos}
\end{figure}

\section{The 200-micron sky}

The 200-$\mu$m (1.5~THz)
atmospheric window is much narrower in wavelength than the more commonly
used sub-millimetre windows, and has much lower transmission.
We modelled this window, along with neighbouring windows, using the 
Cardiff-developed software package FETCH (Araujo et al., 2001;
Hayton et al., 2005).
This is a fast, highly flexible
line-by-line, layer-by-layer, radiative transfer model that uses the
latest High-resolution Transmission (HITRAN) 
molecular spectroscopic database (2004 version)
to calculate the atmospheric transmission as a function of frequency 
(Rothman et al., 2003; 2005). We found that the window extends from 
1428 to 1588~GHz (189--210~$\mu$m), 
with peak transmission at 1486~GHz (202~$\mu$m).
The reason why the peak wavelength is not centred in the window is that
there is an absorbtion line directly in the centre of the window.

\begin{figure}
\setlength{\unitlength}{1mm}
\noindent
\begin{picture}(70,60)
\put(-16,-43){\includegraphics{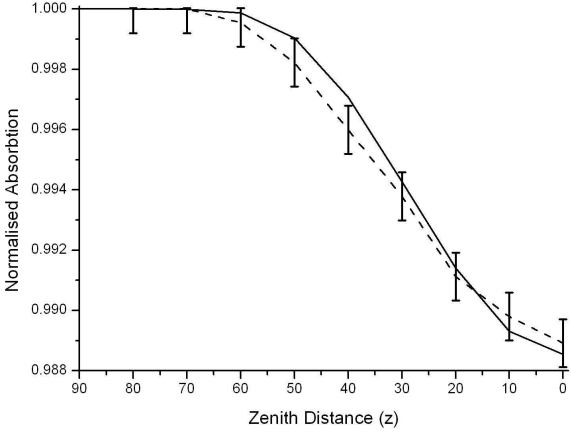}}
\end{picture}
\caption{Plot of a typical set of sky-dip data. Normalised emission
is plotted against zenith distance. The data-points are shown, along
with estimated error-bars, joined by a dashed line.
The solid line is a fit to the data used to obtain $\tau_{1.5THz}$.
See text for details.}
\label{skydip}
\end{figure}

Figure~\ref{atmos} shows the result of the detailed model calculations of
the transmission across the whole of this part of the spectrum, assuming
that the atmospheric precipitable water vapour (PWV) content is 0.5~mm.
The more familiar window at 850~GHz (350~$\mu$m) can
be seen, along with the small transmission window just opening at 1.5~THz
(200~$\mu$m). The peak transmission is low, but nonetheless useable
for part of the time. This calculation is consistent with the measurements
of Paine et al. (2000), under better atmospheric conditions in Chajnantor,
Chile -- see their figure 3.

On the occasions when the sky was totally opaque we measured the relative
responsivity of the seven channels by measuring the sky emission at zenith.
This is essentially a flat, extended source, and allowed us to make a
flat-field of our focal plane array. This remained remarkably constant, 
and was also consistent with our sensitivity measurements made in the
laboratory.

During the times when we could carry out astronomical observations
we calibrated the sky transmission using the method of sky-dipping. This
entails measuring the sky emission at several positions between zenith
and horizon, and modelling the profile as a function of zenith distance, z.
The emission is predicted to have the form $(1 - e^{-\tau sec z})$, and
hence a fitting routine can be used to
calculate the optical depth, $\tau$.

Figure~\ref{skydip} shows the result of a typical sky-dip. On the x-axis
is plotted zenith distance z from 90$^\circ$ to 0$^\circ$, while the
y-axis gives the emission measured from the sky at a given value of z,
normalised to the peak emission at 90$^\circ$. The data-points with
error-bars are shown connected by a dashed line, while 
the solid line shows the model fit to the data for an optimised
value of the optical depth, $\tau$.

Throughout our observations, the JCMT water vapour monitor (WVM) measured
the atmospheric opacity at 183~GHz and from this calculated the opacity at
225~GHz (for historical reasons) in the standard way in which it records
these data (e.g. Archibald et al., 2002). The WVM
operates via a pick-off mirror
at the edge of the JCMT field of view. It updates its estimate of the 
225-GHz opacity every 1.2 seconds. Consequently, at the time of each sky-dip
we know the value of the 225-GHz opacity, so we can plot it against our
measured 1.5-THz opacity.

\begin{figure}
\setlength{\unitlength}{1mm}
\noindent
\begin{picture}(70,40)
\put(-16,-43){\includegraphics{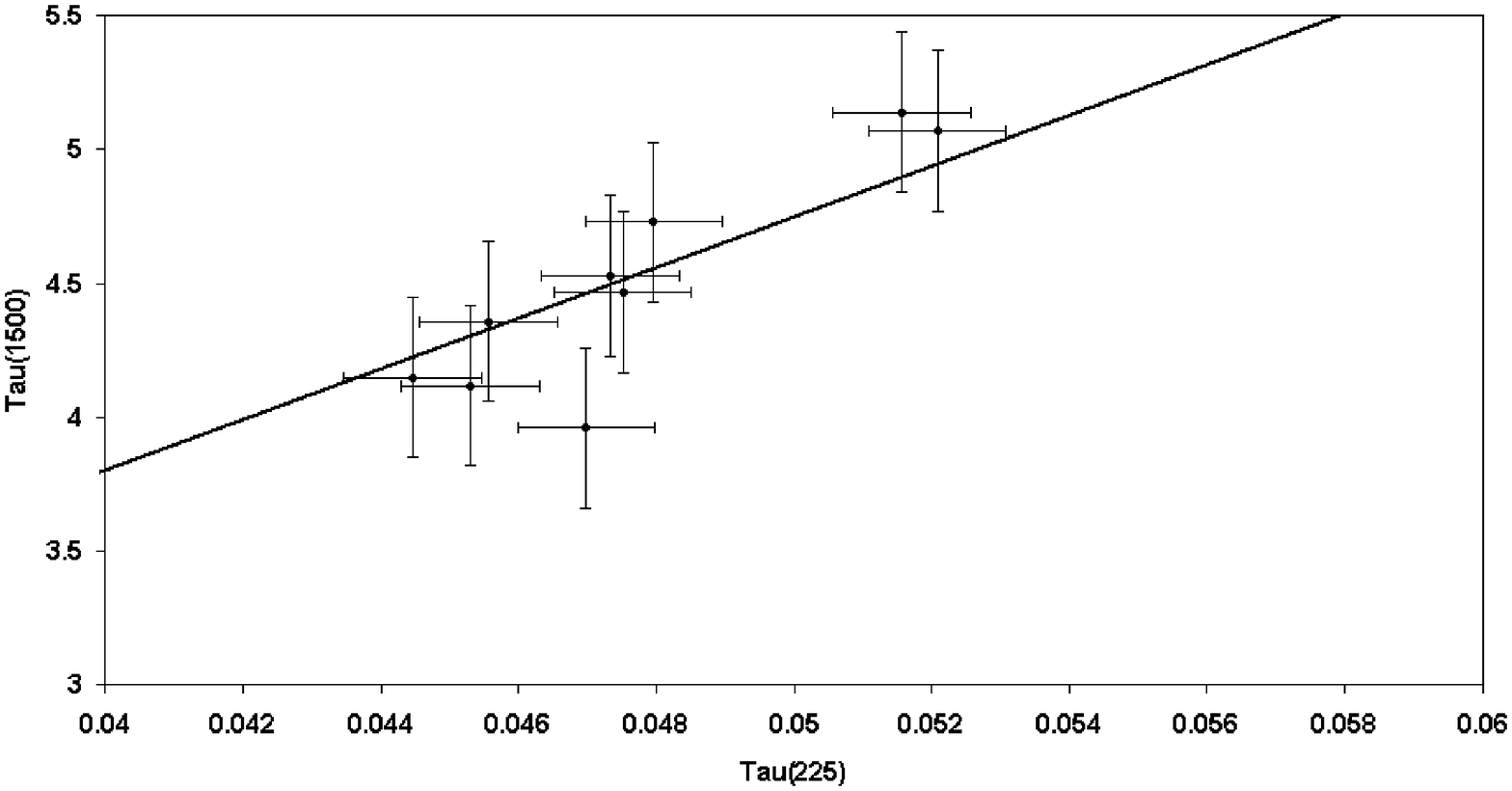}}
\end{picture}
\caption{Plot of $\tau_{1.5THz}$ versus
$\tau_{225GHz}$. The data are shown as crosses. The solid line is a 
least-squares fit to the data. The best fit straight line that we find is
$\tau_{1.5{\rm THz}} = (95 \pm 10) \times \tau_{225{\rm GHz}}$.}
\label{taufit}
\end{figure}

Figure~\ref{taufit} shows the resultant plot of $\tau_{1.5THz}$ versus
$\tau_{225GHz}$. The data are shown as crosses. Our sky model calculations
(see above) predict a linear relation between these two frequencies
over the observed range of values.
The solid line is the resulting least-squares fit to the data.
The best fit straight line that we find is:

\[
\tau_{1.5{\rm THz}} = (95 \pm 10) \times \tau_{225{\rm GHz}} .
\]

\noindent
This result is consistent with our model predictions and
can also be compared with those obtained by previous workers. For
example, Matsushita et al. (1999), found a relation of the form
$\tau_{1.5{\rm THz}} = (105 \pm 32) \times \tau_{225{\rm GHz}}$ for
the observing site at Pampa la Bola in Chile.

These latter authors used an FTS to 
measure the atmospheric opacity across the different wavebands, whereas 
we are comparing our data to WVM data. Nevertheless, recent work shows that
provided all relevant effects are taken into consideration, the results
from FTS and WVM measurements generally agree well (Pardo et al. 2004).
Hence our result is fully consistent with previous work. Therefore we can
use this relation to calibrate our subsequent measurements.

\section{Planet maps}

\subsection{Jupiter}

\begin{figure}
\setlength{\unitlength}{1mm}
\noindent
\begin{picture}(70,60)
\put(-16,-43){\includegraphics{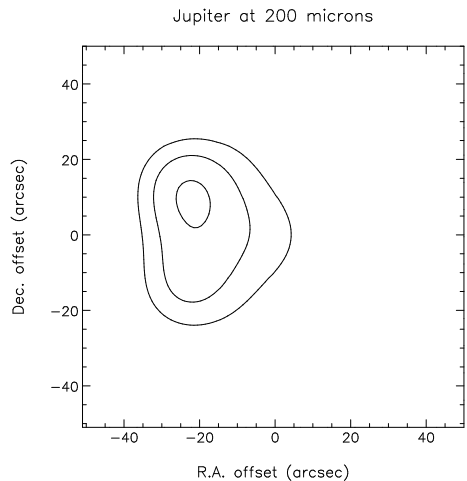}}
\end{picture}
\caption{First-light astronomical image taken with THUMPER: A
200-$\mu$m isophotal contour map of Jupiter.
The field of this image roughly corresponds to the size of the THUMPER
array. The centre of the array is at (0,0). Contour levels are at
50, 70 \& 90\% of peak. The fact that the image appears slightly
non-circular shows that part of the error-beam of the telescope 
may be affecting the data. The optical alignment is also slightly offset.}
\label{jupiter}
\end{figure}

The first astronomical object we imaged was Jupiter.
Figure~\ref{jupiter} shows our 200-$\mu$m map of Jupiter. It was
constructed from two consecutive maps of Jupiter, taken immediately one 
after the other, over the airmass range 1.25 to 1.27.
The two maps both show the same structure, and so they
were co-added to increase the signal-to-noise ratio. 
We note that the source is not exactly centred in the image,
being misaligned by $\sim$30~arcsec. The
telescope absolute pointing accuracy
(based on measurements made with other
instruments) at this time was $\leq$3~arcsec. This
implies that our system alignment was
not perfect. This is not surprising given that this was the first time 
that THUMPER had been mounted on JCMT and this was the first-light
astronomical image.

The flux density
of Jupiter at this wavelength can be predicted from planetary modelling
(Griffin et al., 1986; Orton et al., 1986; Griffin \& Orton 1993).
Based on this, we expect the peak flux density of Jupiter to be 34 kJy/beam.
During the observations the measured value of $\tau_{225 GHz}$ was 0.0625.
Using our relation above, between 1.5THz and 225GHz, this corresponds to
a value of $\tau_{1.5 THz}$ of $\sim$5.95. 

Therefore the observed flux density
at the telescope is predicted to be 21~Jy/beam. The total integration
time per point of the two co-added maps was 100 seconds. The peak
was detected at a level of 
$\sim$3.3~$\sigma$. Therefore we calculate from the Jupiter data that the
noise equivalent flux density (NEFD) of the JCMT-THUMPER combination
is $\sim$63$\pm$10~Jy (1$\sigma$ 1s).

We note that our detection is not at the 5-$\sigma$ level. However, we
believe it is a real detection for a number of reasons: the source was
seen in several pixels simultaneously; the structure in the two maps
that we took was the same; the source appeared in the same place in both 
maps; and the NEFD we calculate from the measurements is consistent with
that predicted from laboratory measurements of the detector system.
In re-gridding the map onto an RA-Dec grid as shown in Figure~\ref{jupiter}
some smoothing naturally occurred, as a pixel scale of 5~arcsec was adopted.

The full-width at half-maximum (FWHM) of Jupiter in our map is 
$\sim$46$\times$38 ($\pm$8) arcsec. At the time of the
observations, the diameter of Jupiter was 42~arcsec. When 
convolved with our 14-arcsec beam, this becomes 44.2~arcsec. Hence our
observations are consistent with this. However, the non-circularity visible
in the map shows that we are also
detecting part of the telescope error beam, which is
predicted to be significant at this wavelength.
We could not trace the error beam further in these data due to their low
signal-to-noise ratio.

It is also possible that if we did not have the instrument exactly in 
focus then this would also contribute to the apparent non-circularity 
of the beam. We focussed the instrument using our model of the telescope
and our calculation of the optimal focal position. However, we did not have
time to check the focus on-source before Jupiter began to set.

\begin{figure}
\setlength{\unitlength}{1mm}
\noindent
\begin{picture}(70,60)
\put(-16,-43){\includegraphics{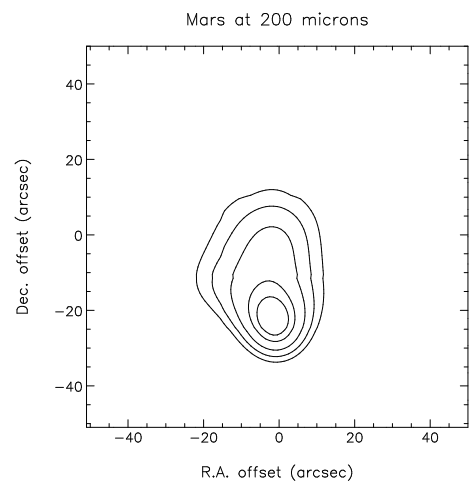}}
\end{picture}
\caption{A 200-$\mu$m isophotal contour map of Mars taken by THUMPER 
on JCMT. The contour levels are at 50, 60, 70, 80 \& 90\% of peak. Once again
the array centre is at (0,0). The alignment is improved in R.A. The extension
to the north is the telescope error beam, which probably also extends
to the south, although this was not mapped.}
\label{mars}
\end{figure}

\subsection{Mars}

Later in the night we imaged Mars as it was rising, from airmass 1.30 to 1.29.
Figure~\ref{mars} shows our map of Mars. Once again the source is not
centred, although we have managed to improve the alignment in R.A.
at least. The flux density
of Mars at this wavelength can be predicted from planetary modelling
in the same way as that of Jupiter quoted above
(Griffin et al., 1986; Orton et al., 1986; Griffin \& Orton 1993).
Based on this, we expect the peak flux density of Mars to be 8.1 kJy/beam.
During the observations the measured value of $\tau_{225 GHz}$ was 0.0455.
Using our relation above, between 1.5~THz and 225~GHz, this corresponds to
a value of $\tau_{1.5 THz}$ of $\sim$4.32.

Therefore the observed flux density
at the telescope is estimated to be 30~Jy/beam. The integration
time per point of the map was 50 seconds. The peak
was detected at a level of 
$\sim$3.2~$\sigma$. Therefore we calculate from the Mars data that the
noise equivalent flux density (NEFD) of the JCMT-THUMPER combination
is $\sim$66$\pm$10~Jy (1$\sigma$ 1s).

We note once again that our detection is not at the 5-$\sigma$ level.
However, we believe it is also
a real detection for similar reasons to those quoted above: the source was
seen in several pixels simultaneously; 
and the NEFD we calculate from the measurements is consistent with
that predicted from laboratory measurements of the detector system
and with that seen in the Jupiter data.

The map of Mars shows a slightly different morphology from that of Jupiter.
The map shows some evidence that the brightest region is more centrally
peaked, and that this central peak sits on an extended plateau.
Conversely, Jupiter is better fitted by a single gaussian.
For Mars the centrally peaked core is of order $\sim$15~arcsec
across, with a more extended lobe to the north. The core is consistent
with the FWHM that would be expected from Mars, since at the time of
the observation the diameter of Mars was 6~arcsec. When convolved
with our beam this produces a gaussian of FWHM 15.2~arcsec, consistent
with the image in Figure~\ref{mars}.

The extended emission to the north is most likely to be the error-beam
of the telescope. The level of the extended
error-lobe is roughly two-thirds of the magnitude of the main beam.
This may also extend to the south and south-east, but
our map does not extend far enough in these directions to say. 
Once again the instrument focus may be adding to the problem. By the
time we repeated the map the shift was coming to an end and the sun had 
risen. Consequently, the conditions worsened (the sky noise level increased)
and we did not detect Mars again in day-time. Our mean NEFD measured on
the planets is therefore $\sim$65$\pm$10~Jy (1$\sigma$ 1s).

\section{The Sun}

Once the Sun had risen we pointed the telescope at the Sun, when it was
in the airmass range 1.44 to 1.43, and detected it
clearly in all channels, at high levels of signal-to-noise ratio of up to
560~$\sigma$. Obviously, we did not map the full extent of the Sun, but
rather used it as a bright, uniform, extended black-body source, and 
chopped across the limb of the Sun. We estimate that the emission from
the Sun is $\sim$1.1~MJy/beam. It was at $\sim$1.4 airmasses during our 
observation, and $\tau_{225GHz}$ was 0.047. We used 5-second integrations, 
so we estimate our NEFD on the Sun to be $\sim$9~Jy (1$\sigma$ 1s).

This is a factor of $\sim$7 better than our NEFD estimated from the 
essentially point-like planet Mars. This implies that only roughly
one-seventh ($\sim$15\%) of the total power incident upon the JCMT dish at
200~$\mu$m is focussed into a 14-arcsec central beam. 
We had predicted that, due to the surface inaccuracies of the JCMT dish
(which was not designed to operate at this high a frequency),
there would be significant power in the side-lobes.

The measurement of the surface accuracy of the dish which is closest in time
to our observations was taken on 2005 February 13, at which time the dish
surface accuracy was found to be 23.8~$\mu$m (Wouterloot 2005).
Using the standard
calculation of Ruze efficiency we estimate that a dish of this accuracy 
would concentrate $\sim$11\% of the total power into a 14-arcsec
beam. This is consistent with the $\sim$15\% we estimate here based on the
ratio of NEFDs measured on Mars and the Sun. Hence we see that all of
our observations are self-consistent.

\section{Conclusions}

We have successfully commissioned the THUMPER camera at JCMT.
We have demonstrated that 200-$\mu$m astronomy is possible from the ground.
We have taken the first ground-based images of Jupiter and Mars at this
wavelength. We have calibrated the 200-$\mu$m sky, and found a relation
between opacities of
$\tau_{1.5{\rm THz}} = (95 \pm 10) \times \tau_{225{\rm GHz}}$.
This is consistent with previous measurements, and also with our modelling
of atmospheric transmission across this section of the electro-magnetic
spectrum. We estimate the NEFD of the JCMT--THUMPER system is $\sim$65~Jy 
(1$\sigma$ 1s).
We find that the error beam of JCMT is substantial at this wavelength,
as predicted, with only $\sim$15\% of the power appearing in the central
14~arcsec. We note that a telescope with a greater surface accuracy
would improve upon this number, and that
other observing sites in even higher or drier 
locations may provide a larger number of useable observing nights
per year. Based on its current NEFD, it seems that
THUMPER could be used to study bright,
high-mass, star-forming regions at higher angular resolution
than has been previously possible at this wavelength.

\section*{Acknowledgments}

The THUMPER Team would like to acknowledge the assistance of the staff
of the JCMT throughout the planning and commissioning of THUMPER. The JCMT
is operated by the Joint Astronomy Centre, Hawaii, on behalf of the UK
Particle Physics and Astronomy Research Council (PPARC), the
Netherlands Organization for Scientific Research (NWO), and the 
Canadian National Research Council (NRC). PPARC are gratefully
acknowledged for grant funding to build THUMPER.

\end{document}